\documentclass[11pt,twoside]{article}
\usepackage{asp2010}

\aspvoltitle{Astronomical Data Analysis Software and Systems XX}
\aspvolauthor{I.~N.~Evans, A.~Accomazzi, D.~J.~Mink and A.~H.~Rots, eds.}
\aspcpryear{2011}
\aspvolume{~}

\resetcounters

\begin{document}

\title{The True Bottleneck of Modern Scientific Computing in Astronomy
}
\author{Igor Chilingarian$^{1,2}$ and Ivan Zolotukhin$^{3,2}$
\affil{$^1$CDS, Observatoire astronomique de Strasbourg, Universit\'e de Strasbourg, CNRS UMR~7550, 11
rue de l'Universit\'e, 67000 Strasbourg, France}
\affil{$^2$Sternberg Astronomical Institute, Moscow State University, 13
Universitetsky prospect, Moscow, 119992, Russia}
\affil{$^3$Observatoire de Paris, LERMA, UMR~8112, 61 Av. de l'Observatoire,
75014 Paris, France}}

\begin{abstract} 
We discuss what hampers the rate of scientific progress in our exponentially
growing world. The rapid increase in technologies leaves the growth of
research result metrics far behind. The reason for this lies
in the education of astronomers lacking basic computer science
aspects crucially important in the data intensive science era.
\end{abstract}

\section{Motivation}

Present-day astronomical instruments and large surveys produce the data flow
increasing exponentially in time. The CPU power required to analyse these
data is also growing with the same pace following the Moore's law; the same
applies to the data storage volume per price unit. However, in astronomy we do not
see the exponential avalanche of scientific results produced with this
computational power. This suggests the presence of \emph{a bottleneck}
somewhere in the loop: \emph{if we consider the system containing three
modules ``A'', ``B'', and ``C'' so that ``A'' is connected to ``C'' via
``B'', then optimizing features in module ``A'' or ``C'' will not produce a
change in the performance of the system until the performance problems in
module ``B'' are addressed.}

Where is the true bottleneck of the scientific computing? Astronomers as
many other scientists, prefer to develop their computational codes and
software systems (including database solutions) themselves often having no
coding skills, insufficient background in algorithms and computational
science.

\section{Code Writing: Astronomers vs Software Engineers}

\subsection{Scientific Software by Scientists}

Most computer programs developed by astronomers without computer science
background, regardless of their purposes (numerical modelling or
simulations, data reduction or visualisation, etc.) often have some specific
common features.

\noindent {\bf (a)}  They are usually written in \emph{Fortran-95}, \emph{-90}, \emph{-77} (or even
prehistoric \emph{Fortran-4} and \emph{-66}). Sometimes high-level
languages (e.g. \emph{IDL}, \emph{MATLAB}) are used. Primitive building 
scripts are used instead of \emph{Makefile}s or more advanced building 
solutions (e.g. \emph{ant}, or \emph{maven} for \emph{Java}). Code is
non-portable.

\noindent {\bf (b)}  They often contain the {\sc goto} statement every 10--20 lines;
names of variables do not follow any conventions, i.e. \emph{a1, a2, aa1};
the code is unreadable: no or bad indentations, very long function bodies
and/or source files. There is a lot of hard-coding of file and device names,
file system paths.

\noindent {\bf (c)}  They are undocumented and full of ``intuitive'' algorithmic solutions,
such as ``re-invented'' sorting and search algorithms, which 
sometimes end up quite far from what computer science students learn at
school.

\noindent {\bf (d)}  The ``multi-layered'' code structure is another typical feature.
When the author is returning to the same program after several months or
years, he/she often finds that the existing procedure/function calls do not
satisfy his/her needs, however is not willing to modify them to keep the
backward compatibility. Then, a wrapper routine is created which is calling
some underlying procedures/functions in a slightly different way. As a
result, after several such periods of development, one can find multiple
(undocumented) interfaces to the same functional blocks.

\noindent {\bf (e)}  {\bf However,} at the end the program does what it is supposed to, because the
author knows exactly what it should do. Even though it may sometimes crash
during run-time or have very poor performance.

\subsection{Scientific Software by IT Engineers}
The software developed by IT engineers in research is notably different.
Here the quality of the final product strongly depends on the job of a 
project manager.

\noindent {\bf (a)}  Usually it is done using a ``real'' programming language:
\emph{C/C++/Java}, primarily because it is virtually impossible to find 
an IT professional developing in \emph{Fortran}.

\noindent {\bf (b)}  All necessary solutions for computational algorithms are conventional 
because the developer at least heard about the ``Art of the computer
programming'' \citep{Knuth78}.

\noindent {\bf (c)}  The code is usually well organized and structured; correct
indentations and variable naming conventions are used; sometimes the author
follow one of the coding styles (e.g. GNU). Therefore, the code becomes readable
and comprehensible.

\noindent {\bf (d)}  The quality and completeness of the documentation strongly depends on
the project manager's competence. It can be from \emph{none} to nearly
perfect.

\noindent {\bf (e)}  {\bf However,} the author often does not understand the physical 
principles behind the algorithm or particular features of the instrumentation 
making the data looking as they are, therefore some bad surprises are possible.
For example, some arithmetic bugs leading to the results which are wrong by
many orders of magnitude from what is expected cannot be spotted by a
software engineer because for him/her it is ``just a number''. This may
dramatically slow down the development.

\subsection{Databases by Scientists}

The worst class of software solutions is probably DBs developed by
researchers.

\noindent {\bf (a)}  Often they contain custom implementation in \emph{Fortran} or \emph{IDL} of 
re-invented indexing solutions and primitive requests to the data. Indices
and data tables are stored in a proprietary undocumented binary format.

\noindent {\bf (b)}  If an existing database management system (DBMS) is used, then the 
DB usually contains one or several flat tables without mutual
links, i.e. no data model.

\noindent {\bf (c)}  DB constraints are not used for consistency checks, in some rare
cases they are implemented externally in a DB management interface
(also often written in \emph{Fortran}).

\noindent {\bf (d)}  User interfaces, both application programming interface (API) and
web front-end are undocumented, have very low usability and terrible design.

\section{Bad and Good Examples}

For obvious reasons we will not cite the corresponding references for bad
examples. The list of good examples is neither exhaustive nor complete.

\subsection{Bad Example \#1: an Unnamed Galaxy Catalogue}
The project is very interesting scientifically and recognised in the
community. But$\dots$

\noindent {\bf (a)}  There is no access interface on the web.

\noindent {\bf (b)}  The data are distributed as a set of dozens of FITS tables with a 
total volume $>$10Gb and \emph{IDL} access routines to perform queries on
these tables. One has to download nearly everything in order to study just 
a handful of objects.

\noindent {\bf (c)}  Therefore, huge memory requirements if one uses the 
whole catalogue at once.

\noindent {\bf (d)}  Therefore, very slow and inefficient data access and selection.

\subsection{Bad Example \#2: an Unnamed Database Using \emph{PostgreSQL}}

\noindent {\bf (a)}  DB administration and ingestion interface (implemented in
\emph{Fortran}) has a function with over 250 arguments

\noindent {\bf (b)}  Inside the DB restore script, to delete a record from a table, 
instead of 

\emph{DELETE FROM table1 WHERE field1=value1} the authors do:

\emph{pg\_dump -t table1 mydb $|$ grep -v value1 $|$ pg\_restore -c mydb}

\noindent {\bf (c)}  One of the stored procedures which is triggered on \emph{INSERT},
connects externally to the same DB and making some selections using
this new connection. Obviously, it cannot see the changes introduced before
the trigger had been fired because the transaction has not been committed.

\subsection{Good Examples \#1: Technologically Advanced Projects}

{\bf 1. HLA} - the Hubble Legacy Archive (\url{http://hla.stsci.edu/}).
Innovative solutions implemented
inside HLA include: (a) Virtual Observatory standard interfaces (Simple
Image Access Protocol) as a hidden middleware; (b) XSLT transformation of
\emph{VOTables} into \emph{AJAX}-enabled HTML pages; (c) advanced
visualisation tools.

\noindent {\bf 2. SDSS CasJobs} (\url{http://cas.sdss.org/CasJobs},
\citealp{Szalay+02}).  Efficient and easy-to-use access to a large DB
featuring user management, user table upload, I/O of tabular data in
different formats, comprehensive SQL query builder.

\noindent
{\bf 3. GalexView} (\url{http://galex.stsci.edu/GalexView/}) -- a Flash-based
interactive web-access to the GALEX satellite images.

\noindent
{\bf 4. Millennium Simulation}
(\url{http://www.mpa-garching.mpg.de/millennium/} by G.~Lemson) -- access to
the DB containing the results of large cosmological simulations
with a comprehensive data model and full SQL access.

\noindent {\bf 5. GalMer} (\url{http://galmer.obspm.fr/},
\citealp{Chilingarian+10}) - a DB to access numerical simulations of merging
and interacting galaxies. The projects implements a set of Virtual
Observatory (VO) standards, features efficient interactive preview
visualisation of the datasets on the server side, complex on-the-fly data
analysis algorithms. The \emph{JavaScript}-powered web-interface working in
most modern browsers is integrated with VO tools in order to visualise
complex datasets \citep{CZ08,ZC08}.

\subsection{Good Examples \#2: Computations, Data Analysis and Visualisation}

{\bf 1. GADGET-2} by V.~\citet{Springel05}, a cosmological simulation code
that is well documented and easily extensible: there are numerous
third-party add-ons implementing different physical phenomena, e.g.
radiative transfer, metallicity evolution in galaxies.

\noindent {\bf 2. SExtractor} \citep{BA96}: a software to perform object
extraction and photometry from CCD images has very intuitive configuration,
although outdated documentation. It is relatively clearly coded

\noindent {\bf 3. TOPCAT/STILTS} \citep{Taylor05,Taylor06} -- the best
available platform independent table manipulation software integrated with 
VO services and resources.

\noindent {\bf 4. CDS Aladin} \citep{Bonnarel+00} -- a VO
data browser for images and catalogues.

\noindent {\bf 5. SAOImage DS9} \citep{JM03} -- probably the most frequently
used desktop FITS visualisation software in astronomy implementing some VO
data access methods.

\section{The Main Message and a Possible Solution}

It turns out that all ``good examples'' were developed either by
professional astronomers with very strong IT/CS background or by IT/CS
professionals working closely with astronomers for years and
understanding astronomy. One cannot simply hire an industrial software
engineer to develop astronomical software and/or an archive and/or a
database.

A possible solution is \emph{to change the teaching paradigm for students in
astronomy}. Basic courses in algorithms, programming, software development
and maintenance have to be made mandatory in the education of modern
astronomers and physicists; advanced courses should be recommended to some
of them. The \emph{Fortran} language is now obsolete and we have to accept
this. Instead of teaching research students to \emph{Fortran} programming,
one should teach how to interface legacy \emph{Fortran} code in
\emph{C/C++}.

As soon as this \emph{bottleneck} is resolved, the avalanche of discoveries
will loom.


\bibliography{O14_2}
\bibliographystyle{asp2010}

\end{document}